\documentclass[sigconf,screen]{acmart}
\renewcommand\footnotetextcopyrightpermission[1]{}
\usepackage{graphicx}    % 插入图片
\usepackage{booktabs}    % 提供 \toprule, \midrule, \bottomrule 三线表支持
\usepackage{multirow}    % 提供多行合并 \multirow
\usepackage{colortbl}    % 提供表格行颜色 \rowcolor 和 \cellcolor
\usepackage{xcolor}      % 颜色支持
\usepackage{pifont}      % 提供 \ding{51} (打勾) 和 \ding{55} (打叉)
\usepackage{tabularx}    % 提供 tabularx 环境 (用于自适应列宽)
\usepackage{caption}     % 控制标题格式
\usepackage{subcaption}  % 提供子图支持 \begin{subfigure} (已去重)

\usepackage{multicol}
\usepackage{tcolorbox}
\definecolor{mygreen}{HTML}{3cb44b}

\usepackage{enumitem}
\usepackage{amsmath}
\usepackage{amsfonts}
\usepackage{mathtools}

\usepackage{amssymb}     % 提供 \times 等数学符号 (已合并至此处)

\usepackage{algorithm}    % 提供 \begin{algorithm} 浮动体环境和 \caption
\usepackage{algorithmic}  % 提供 \REQUIRE, \STATE, \FOR, \ENDFOR 等大写控制命令

\usepackage{bm}           % 允许使用 \bm{} 加粗斜体希腊字母
\usepackage{dsfont}       % 用于输入更美观的双线空心字母（如指示函数 \mathds{1}）

\usepackage[table]{xcolor}
\definecolor{mygray}{gray}{0.92}
\newcolumntype{L}[1]{>{\raggedright\let\newline\\\arraybackslash\hspace{0pt}}p{#1}}

\definecolor{mygray}{gray}{0.92} % 定义淡灰色
\newcolumntype{Y}{>{\centering\arraybackslash}X}

\AtBeginDocument{%
  }

\setcopyright{acmlicensed}
\copyrightyear{2018}
\acmYear{2018}
\acmDOI{XXXXXXX.XXXXXXX}
\acmConference[Conference acronym 'XX]{Make sure to enter the correct
  conference title from your rights confirmation email}{June 03--05,
  2018}{Woodstock, NY}
\acmISBN{978-1-4503-XXXX-X/2018/06}

\begin{document}

%%
%% The "title" command has an optional parameter,
%% allowing the author to define a "short title" to be used in page headers.
\title{FedTaste: Topology-Aware Structural Transfer for Multimodal Federated Learning with Missing Modalities}

%%
%% The "author" command and its associated commands are used to define
%% the authors and their affiliations.
%% Of note is the shared affiliation of the first two authors, and the
%% "authornote" and "authornotemark" commands
%% used to denote shared contribution to the research.
\author{Haochen Liang}
\email{lianghc@g.ecc.u-tokyo.ac.jp}
\affiliation{%
  \institution{The University of Tokyo}
  \city{Tokyo}
  \country{Japan}
}

\author{Jie Zhang}
\email{jz@stu.cqut.edu.cn}
\affiliation{%
  \institution{Great Bay University}
  \city{Dongguan}
  \country{China}
}

\author{Hideya Ochiai}
\email{ochiai@g.ecc.u-tokyo.ac.jp}
\affiliation{%
  \institution{The University of Tokyo}
  \city{Tokyo}
  \country{Japan}
}
%%
%% By default, the full list of authors will be used in the page
%% headers. Often, this list is too long, and will overlap
%% other information printed in the page headers. This command allows
%% the author to define a more concise list
%% of authors' names for this purpose.

%%
%% The abstract is a short summary of the work to be presented in the
%% article.
\begin{abstract}
Multimodal Federated Learning is often challenged by arbitrary modality missingness and Non-IID data distributions, which lead to severe representation drift and hinder effective collaboration across clients. Existing methods typically rely on generative imputation, external auxiliary data, or isolated unimodal training to bridge modality gaps, often incurring substantial communication and computational costs as well as potential privacy risks. To address these limitations, we propose FedTaste, a parameter-efficient framework for topology-aware structural transfer in Multimodal Federated Learning with missing modalities. Instead of aligning fragile first-order features, FedTaste focuses on more stable group-level semantic relations. Specifically, FedTaste leverages frozen foundation models to extract a joint multimodal topology from full-modality clients, which is then consolidated by the server into a global structural blueprint. To adapt clients with missing modalities, we introduce Modality-Adaptive Structural Prompts together with spectral consistency regularization, enabling lightweight branch-specific adaptation that aligns local partial representations with the shared blueprint. In this way, FedTaste avoids explicit modality imputation while preserving shared semantic structure across clients. Extensive experiments demonstrate that FedTaste consistently achieves superior performance across multiple datasets and challenging Non-IID settings, while substantially reducing communication overhead compared with existing methods.
\end{abstract}

%%
%% The code below is generated by the tool at http://dl.acm.org/ccs.cfm.
%% Please copy and paste the code instead of the example below.
%%


%%
%% Keywords. The author(s) should pick words that accurately describe
%% the work being presented. Separate the keywords with commas.
%% A "teaser" image appears between the author and affiliation
%% information and the body of the document, and typically spans the
%% page.

%%
%% This command processes the author and affiliation and title
%% information and builds the first part of the formatted document.
\maketitle

\begin{figure}[t]
    \centering
    \includegraphics[width=0.98\linewidth]{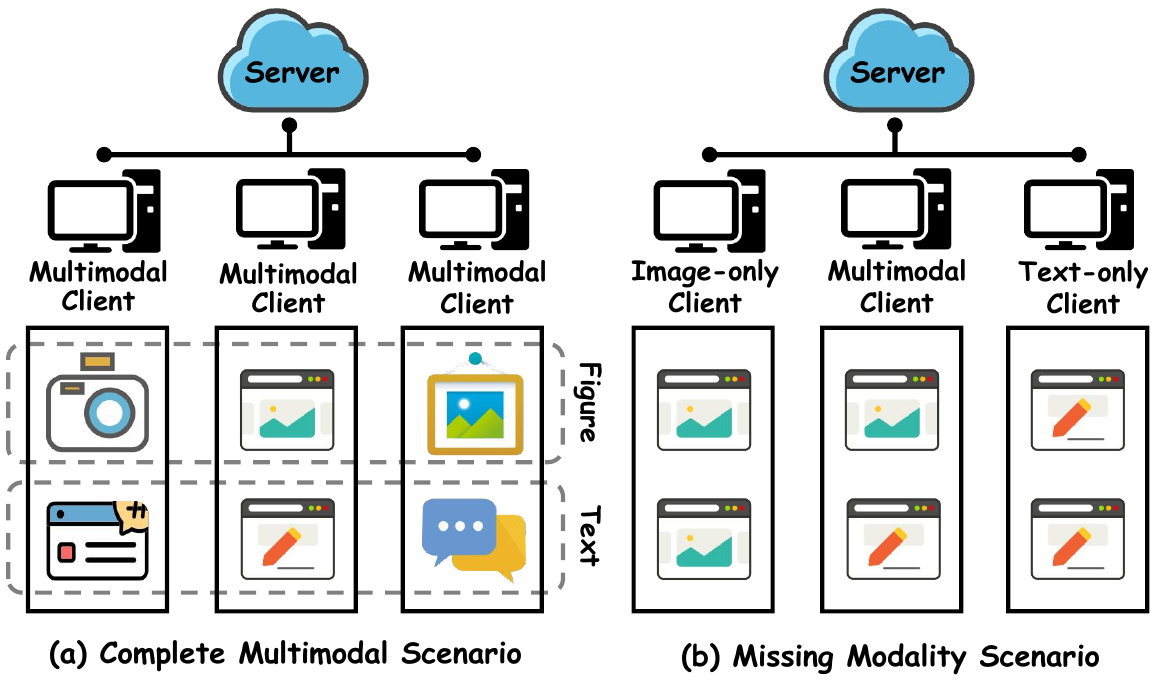}
    \vspace{-0.5em}
    \caption{Two major challenges in federated multimodal learning. \textbf{(a) Data heterogeneity}: clients have complete modalities, but their local data distributions vary substantially across sources. \textbf{(b) Modality heterogeneity}: in more realistic settings, clients may entirely lack certain modalities.}
    \label{fig1}
    \vspace{-1em}
\end{figure}

\section{Introduction}

Multimodal Federated Learning (MFL)~\cite{che2023multimodal} enables multiple decentralized clients to collaboratively train models over heterogeneous data sources while preserving data privacy. By exploiting complementary information across diverse modalities, MFL has shown promise in applications including autonomous driving~\cite{li2021privacy}, healthcare informatics~\cite{nguyen2022federated}, sensors~\cite{bao2025robust} and cross-modal retrieval~\cite{zong2021fedcmr}. Its goal is to learn a unified representation space in which semantically related features from different modalities are robustly aligned. In practice, however, MFL is fundamentally challenged by arbitrary modality missingness and Non-IID data distributions~\cite{huang2024multimodal}. Hardware heterogeneity, unstable sensor availability, and privacy constraints cause modality scarcity across clients~\cite{feng2024robust}. When this incompleteness is coupled with severe local heterogeneity, cross-modal alignment becomes ill-posed. Without complete multimodal pairs as structural anchors, local representation spaces diverge, leading to representation drift and degraded global performance~\cite{song2024tackling}.

\begin{figure*}[t]
    \centering
    \includegraphics[width=0.98\textwidth]{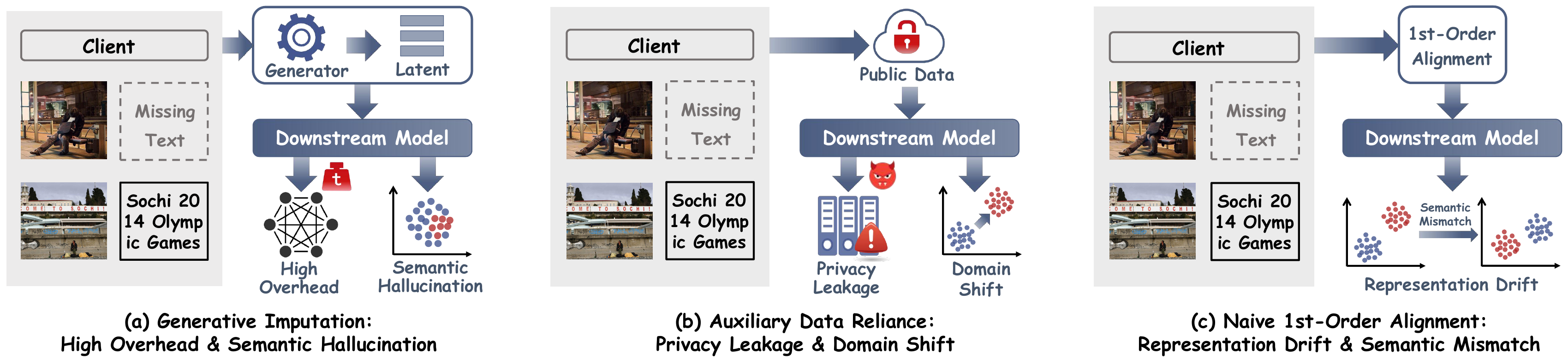}
    \vspace{-0.5em}
    \caption{Motivation of the FedTaste framework. Existing paradigms for missing-modality MFL suffer from three limitations: \textbf{(a)} generative imputation introduces semantic hallucination and high overhead; \textbf{(b)} reliance on auxiliary data is vulnerable to domain shift and privacy risks; \textbf{(c)} naive first-order alignment breaks down under Non-IID settings due to representation drift.}
    \label{fig:motivation}
    \vspace{-0.5em}
\end{figure*}

Existing paradigms for handling modality gaps in federated settings mainly rely on generative imputation~\cite{cao2026heterogeneous}, auxiliary data~\cite{yu2023multimodal}, or unimodal training~\cite{sun2024towards}. While effective in specific scenarios, these strategies often suffer from limited robustness and efficiency. Generative approaches typically introduce considerable computational overhead, making them impractical for resource-constrained edge devices, and may inject semantic hallucinations through unstable reconstruction~\cite{liang2026fedcoop}. Methods based on exogenous priors or public datasets are additionally exposed to domain shifts and privacy vulnerabilities, compromising the closed-loop nature of federated systems~\cite{chen2024fedmbridge}. Unimodal aggregation, in turn, fails to preserve cross-modal semantic associations~\cite{zhang2025unimodal}. These intrinsic limitations call for a lightweight, robust, and privacy-preserving mechanism for multimodal alignment in federated networks.

Recent vision-language foundation models, especially CLIP~\cite{radford2021learning}, provide strong semantic priors for multimodal learning~\cite{shi2024clip}. Yet integrating such models into MFL remains difficult when modality incompleteness is severe. Most existing methods still rely on first-order point-wise alignment, namely matching absolute coordinates in the embedding space~\cite{feng2023fedmultimodal}. Under severe Non-IID conditions, however, local embedding spaces can drift substantially and asynchronously across clients~\cite{tan2023fedsea}. This geometric distortion makes coordinate-level absolute alignment fragile, unreliable, and prone to negative transfer~\cite{chen2022towards}. This suggests that effective federated multimodal alignment should rely less on absolute coordinates and more on structural relations that remain stable under representation drift~\cite{jiang2022harmofl}. Specifically, rather than relying on absolute coordinates, utilizing relative distances among semantic anchors provides a stable second-order structure that naturally resists the severe spatial shifts caused by heterogeneous local data~\cite{mendieta2022local}.

Motivated by this, we propose FedTaste, a topology-aware structural transfer framework for MFL. The key insight is that, while absolute feature coordinates are highly sensitive to drift~\cite{wu2024topology}, second-order semantic relations among group-level anchors remain more stable across clients and modalities~\cite{hu2026fedtopo}. FedTaste extracts a Joint Multimodal Topology from modality-complete teacher clients and consolidates it on the server into a confidence-aware global structural blueprint. Missing-modality clients are then adapted through Modality-Adaptive Structural Prompts (MASP) with spectral consistency regularization. FedTaste further adopts an asymmetric federated optimization strategy, where teacher clients upload compact structural payloads and student clients update only lightweight prompts. Extensive experiments show that FedTaste consistently outperforms existing methods under challenging Non-IID settings with lower communication overhead.

Our contributions are summarized as follows:
\begin{itemize}
    \item We present FedTaste, a topology-aware structural transfer framework for multimodal alignment in federated settings that addresses modality missingness and representation drift through stable second-order semantic relations.
    \item We introduce Modality-Adaptive Structural Prompts (MASP) with spectral consistency regularization for lightweight missing-modality adaptation. Coupled with an asymmetric federated optimization strategy, it avoids full-model updating and costly generative imputation.
    \item We develop a confidence-aware global structural blueprint that distills a Joint Multimodal Topology from modality-complete participants and provides a stable, noise-reduced structural target for missing-modality adaptation.
    \item Extensive experiments on multiple benchmark datasets demonstrate the strong performance and communication efficiency of FedTaste compared with existing methods.
\end{itemize}

\begin{figure*}[t]
    \centering
    \includegraphics[width=0.99\textwidth]{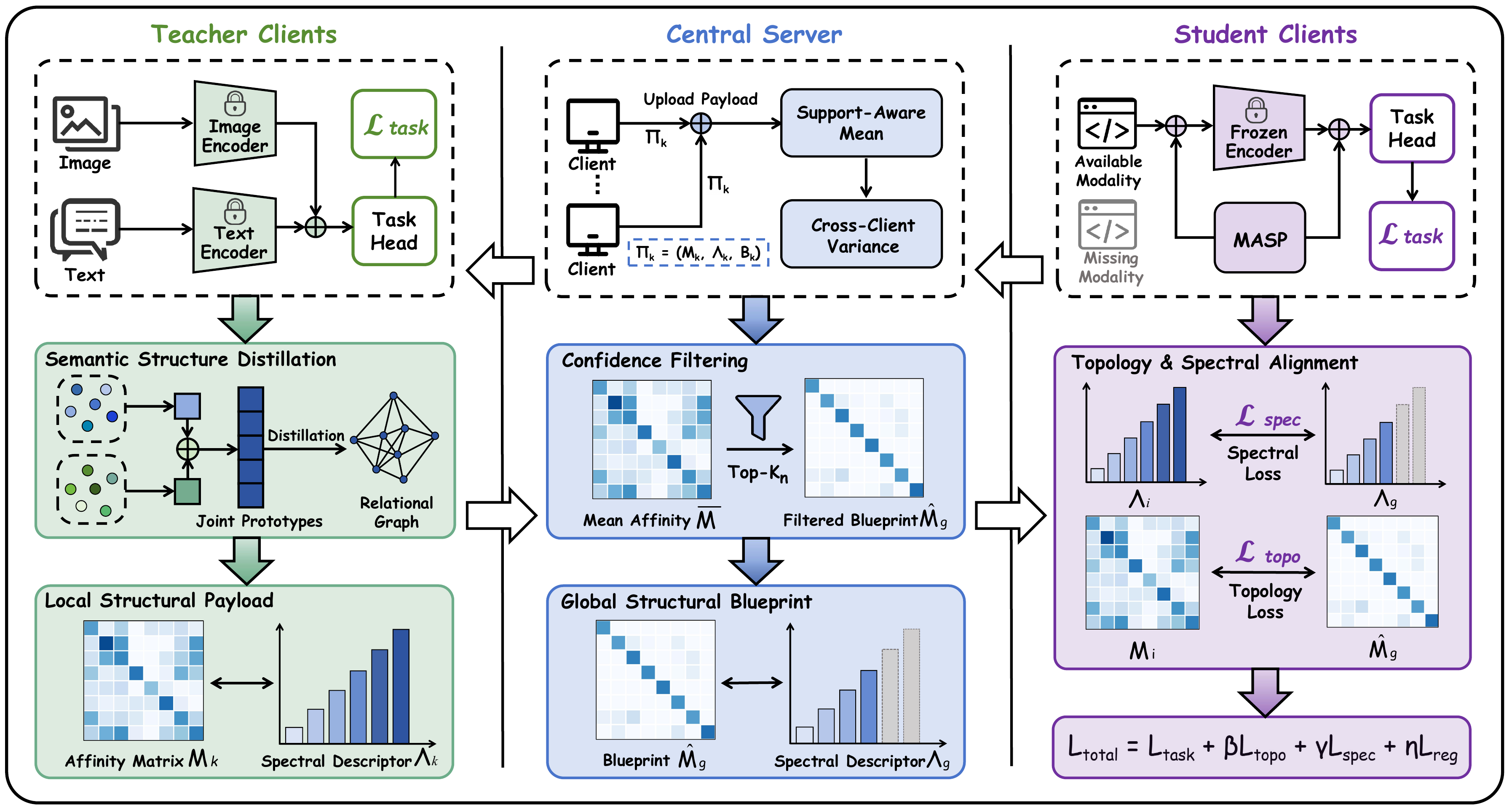}
    \caption{
    Overview of the FedTaste framework. FedTaste addresses federated multimodal learning with missing modalities through three stages: semantic structure distillation on full-modality clients, confidence-aware global blueprint construction on the server, and lightweight topology-guided adaptation on missing-modality clients. Full-modality clients extract multimodal structural payloads, the server consolidates them into a global structural blueprint, and missing-modality clients use Modality-Adaptive Structural Prompts to align local partial representations under shared topology and spectral guidance.
    }
    \label{fig4}
\end{figure*}

\section{Related Work}

\subsection{Missing Modalities in MFL}

Existing strategies for handling modality missingness in federated learning can be broadly grouped into three categories based on their cross-modal alignment mechanisms.

The first category addresses modality incompleteness via generative imputation. Representative methods, such as FL-MRSG~\cite{cao2026heterogeneous} and FedMobile~\cite{liu2025fedmobile}, utilize auxiliary generative models or knowledge-distillation-based feature generators to synthesize absent features. While effective, these approaches~\cite{poudel2025multimodal} impose severe computational burdens on resource-constrained edge devices and risk introducing semantically inconsistent pseudo-features when local data distributions are highly heterogeneous~\cite{yan2025federated}.

A second line of research leverages external public data to guide alignment. For instance, CreamFL~\cite{yu2023multimodal} employs auxiliary datasets to transfer knowledge across clients with varying modality profiles. Similarly, recent methods utilize public datasets either for retrieval-based modality augmentation (e.g., CAR-MFL~\cite{poudel2024car}) or to extract representations for server-side knowledge distillation (e.g., FedAFD~\cite{tan2026fedafd}). Although these strategies improve cross-client consistency, relying on exogenous data inevitably introduces domain shift risks and may weaken the closed-loop or privacy-preserving assumptions of federated systems~\cite{hsu2024federated}.

The third category focuses on direct feature or prototype alignment, bypassing explicit generation. Methods like FedCola~\cite{sun2024towards} and FedIoT~\cite{zhao2022multimodal} align or aggregate available local representations to bridge intra-modal and cross-modal gaps. While computationally lighter, their dependence on first-order absolute coordinate matching renders them highly susceptible to representation drift, particularly when local embedding spaces diverge under severe Non-IID settings~\cite{yu2026robust}.

\subsection{Relational and Structural Learning}

Rather than focusing on absolute feature coordinates, relational learning emphasizes the structural topology among data points or semantic centers~\cite{chen2024enhancing}. This concept is widely adopted in centralized paradigms, such as relational knowledge distillation~\cite{yang2022geometric} and graph-based domain adaptation~\cite{ngo2025higda}, where second-order similarities facilitate structural knowledge transfer. Theoretical studies indicate that second-order relational structure is often more stable than absolute coordinates under representation shifts, offering a robust foundation against spatial distortions~\cite{wu2024topology}.

Despite these advantages, the application of structural alignment in MFL remains underexplored, especially under arbitrary modality missingness and severe Non-IID heterogeneity~\cite{kavalionak2021impact}. Conventional federated approaches predominantly rely on point-wise feature matching, which easily collapses under these compounded challenges~\cite{wei2024joint}. FedTaste bridges this literature gap by introducing topology-aware structural transfer into federated networks. This paradigm shift—from absolute coordinate matching to second-order geometric alignment—motivates our design to empower missing-modality clients to reliably anchor their partial representations to a shared relational blueprint~\cite{wang2024fedmmr}. Distilling a globally consistent semantic topology rather than isolated features calibrates local representation spaces without the computational overhead of generative models or the domain mismatch risks of external datasets~\cite{lin2023federated}.

\section{Methodology}
\label{sec:methodology}

\subsection{Problem Setup}
\label{subsec:problem_setup}

We study multimodal federated learning under arbitrary modality missingness. Let $\mathcal{C}=\{1,\dots,N\}$ denote the set of clients. Each client $i$ owns a local dataset $\mathcal{D}_i$ and a modality subset $\mathcal{M}_i \subseteq \{v,l\}$, where $v$ and $l$ denote the visual and language modalities, respectively. Clients with $\mathcal{M}_i=\{v,l\}$ are treated as full-modality teachers and form the set $\mathcal{C}_T$, while clients with $\mathcal{M}_i=\{v\}$ or $\mathcal{M}_i=\{l\}$ are treated as missing-modality students and form the set $\mathcal{C}_S$.

To make the formulation applicable to both retrieval and labeled scenarios, we introduce a unified set of semantic groups $\mathcal{G}=\{1,\dots,G\}$ as the basic structural units. In retrieval benchmarks, a group corresponds to an image-centered pairing unit or its associated semantic anchor; in labeled settings, it degenerates to the class index. Under this formulation, the objective of FedTaste is not to reconstruct absent modalities explicitly. Instead, it aims to transfer stable group-level semantic structure from $\mathcal{C}_T$ to $\mathcal{C}_S$ under severe Non-IID heterogeneity, while keeping student-side adaptation lightweight.

\subsection{Overview}
\label{subsec:overview}

FedTaste is built on the observation that under severe Non-IID conditions, absolute feature coordinates are often unstable across clients, whereas relative semantic relations among groups are substantially more transferable. Consequently, directly aligning first-order coordinates can be fragile, especially when local representations drift due to statistical heterogeneity and modality incompleteness. FedTaste therefore shifts the target of cross-modal alignment from raw features to \emph{group-level relational structures}.

The framework is organized around two coupled channels. The \emph{structure channel} distills a relational graph from full-modality teachers, consolidates it on the server into a confidence-aware global structural blueprint, and broadcasts this blueprint to missing-modality students. The \emph{parameter channel} enforces parameter-efficient adaptation: to preserve the semantic prior of the foundation models, all clients keep the shared pretrained multimodal backbone frozen and update only a lightweight prompt subset. The asymmetric federated design thus lies purely in their structural roles: teachers generate and upload multimodal structural payloads, which are consolidated by the server into a global blueprint, while students receive and adapt to it without relying on heavy generative imputation or external auxiliary data.

\subsection{Teacher-Side Semantic Structure Distillation}
\label{subsec:teacher_distillation}

For each participating teacher client $k \in \mathcal{C}_T$, FedTaste adopts the pre-trained visual and text encoders of CLIP (ViT-B/32). While keeping these backbones frozen, it optimizes local Modality-Adaptive Structural Prompts (MASP) and then extracts a compact structural summary. Let $E_v$ and $E_l$ denote the prompt-conditioned visual and language encoders. For each semantic group $g \in \mathcal{G}$, the teacher computes modality-specific group prototypes by averaging normalized features within the group:
\begin{equation}
p_{m,k}^{g}=\frac{1}{|\mathcal{D}_{k,m}^{g}|}\sum_{x \in \mathcal{D}_{k,m}^{g}} \frac{E_m(x)}{\|E_m(x)\|_2}, \qquad m \in \{v,l\},
\end{equation}
where $\mathcal{D}_{k,m}^{g}$ denotes the subset of modality-$m$ samples from group $g$ stored on client $k$. Since teacher clients observe both modalities, the two prototypes are fused into a joint semantic anchor:
\begin{equation}
q_k^{g}=\frac{p_{v,k}^{g}+p_{l,k}^{g}}{\|p_{v,k}^{g}+p_{l,k}^{g}\|_2}.
\end{equation}
This group-level representation is generally more stable than raw sample features and provides a natural basis for building a local semantic graph.

Based on these anchors, client $k$ constructs a local relational graph over groups. Let $B_k \in \{0,1\}^{G \times G}$ be the support mask, where $(B_k)_{gh}=1$ indicates that both groups $g$ and $h$ are locally observed. The affinity between two supported groups is defined as
\begin{equation}
m_{gh}^{(k)}=
\begin{cases}
\exp \left(\dfrac{\langle q_k^{g}, q_k^{h} \rangle}{\tau_s}\right), & g \neq h \ \text{and}\ (B_k)_{gh}=1,\\[6pt]
0, & \text{otherwise},
\end{cases}
\end{equation}
which yields a non-negative affinity matrix $M_k \in \mathbb{R}^{G \times G}$. Unlike direct feature matching, this matrix captures second-order relations among semantic groups and is therefore less sensitive to client-specific coordinate drift.

To summarize the macro-level structural properties of $M_k$, we compute its symmetric normalized Laplacian
\begin{equation}
L_k=I-D_k^{-1/2}M_kD_k^{-1/2},
\end{equation}
where $(D_k)_{gg}=\sum_h m_{gh}^{(k)}$. We then retain the smallest $K$ non-trivial eigenvalues of $L_k$ as the teacher-side spectral descriptor:
\begin{equation}
\Lambda_k=\left[\lambda_2^{(k)},\lambda_3^{(k)},\dots,\lambda_{K+1}^{(k)}\right]^\top.
\end{equation}
Here, $M_k$ captures fine-grained pairwise affinities, while $\Lambda_k$ characterizes the low-frequency properties of the local semantic graph. Using the normalized Laplacian bounds the eigenvalues and inherently improves practical stability under heterogeneous local supports. The structural payload uploaded by teacher client $k$ is therefore $\Pi_k=(M_k,\Lambda_k,B_k)$.

\subsection{Server-Side Global Structural Blueprint}
\label{subsec:server_blueprint}

Teacher graphs extracted under severe Non-IID distributions inevitably contain incomplete support and client-specific noise. FedTaste therefore does not aggregate all local edges indiscriminately. Instead, the server consolidates teacher payloads into a \emph{confidence-aware global structural blueprint} that emphasizes relations that are both sufficiently supported and cross-client stable.

For each edge $(g,h)$, the server first counts its support frequency as $n_{gh}=\sum_{k \in \mathcal{C}_T}(B_k)_{gh}$. It then computes the support-aware mean affinity
$\bar{m}_{gh}=\sum_{k \in \mathcal{C}_T}(B_k)_{gh}m_{gh}^{(k)}/\max(n_{gh},1)$
and the corresponding cross-client variance
$\sigma_{gh}^2=\sum_{k \in \mathcal{C}_T}(B_k)_{gh}(m_{gh}^{(k)}-\bar{m}_{gh})^2/\max(n_{gh},1)$.
Collecting all $\bar{m}_{gh}$ yields the mean affinity matrix $\bar{M} \in \mathbb{R}^{G \times G}$. 

To preserve only globally reliable relations, the server constructs a confidence mask $S \in \{0,1\}^{G \times G}$ by retaining, for each row, the top-$K_n$ strongest edges among those satisfying a variance threshold $\rho$:
\begin{equation}
S_{gh}=\mathbb{I}[n_{gh}>0]\cdot \mathbb{I}[\sigma_{gh}^2 \le \rho]\cdot \mathbb{I}\!\left[h \in \operatorname{TopK}_g(\bar{M},K_n)\right].
\end{equation}
The cross-client variance $\sigma_{gh}^2$ acts as a crucial indicator of structural consensus; relations exhibiting high variance are typically artifacts of local data skew rather than true semantic affinities. The resulting global structural blueprint is represented as
\begin{equation}
\hat{M}_{\text{g}}=S \odot \bar{M}.
\end{equation}
This filtering step ensures that FedTaste explicitly removes weakly supported or unstable edges, yielding a blueprint that reliably reflects transferable semantic structure at the federation level.

In parallel, teacher-side spectral summaries are aggregated by simple averaging:
\begin{equation}
\Lambda_{\text{g}}=\frac{1}{|\mathcal{C}_T|}\sum_{k \in \mathcal{C}_T}\Lambda_k.
\end{equation}
The server also merges teacher support into a global activity mask
$A_{\text{g}}=\mathbb{I}\!\left[\sum_{k \in \mathcal{C}_T} B_k > 0\right]$.
The final blueprint broadcast to students is therefore $\mathcal{B}=(\hat{M}_{\text{g}},S,A_{\text{g}},\Lambda_{\text{g}})$, which provides both edge-level and spectral guidance.

\subsection{Student-Side Lightweight Structural Adaptation}
\label{subsec:student_adaptation}

Following the parameter-efficient protocol, student clients also keep the pretrained backbone frozen and optimize only their Modality-Adaptive Structural Prompts (MASP). For each student client $i \in \mathcal{C}_S$, let $a_i \in \{v,l\}$ denote its observed modality. We attach a set of learnable prompts $P_i=\{P_i^{(d)}\}_d$ to the active encoder stream. The resulting feature for input $x$ is
\begin{equation}
f_i(x)=\frac{\tilde{E}_{a_i}(x;P_i)}{\|\tilde{E}_{a_i}(x;P_i)\|_2},
\end{equation}
where $\tilde{E}_{a_i}$ denotes the prompt-conditioned active branch. This design preserves the transferable prior of the multimodal encoder while allowing students to adjust their local structure through a minimal parameter footprint.

\paragraph{Local task objective.}
Since student clients observe only one modality, the task loss must be defined entirely within the available branch. Given a mini-batch $\mathcal{B}_i=\{(x_n,y_n)\}_{n=1}^{B}$, where group labels $y_n \in \mathcal{G}$ strictly correspond to the unified semantic grouping defined in Section~\ref{subsec:problem_setup}, we use a group-supervised contrastive objective. Let $\mathcal{P}(n)=\{m \neq n \mid y_m=y_n\}$ denote the positive set for sample $n$. The task loss is defined as
\begin{equation}
\mathcal{L}_{\text{task}}^{(i)}=
-\frac{1}{B}\sum_{n=1}^{B}
\frac{1}{|\mathcal{P}(n)|}
\sum_{p \in \mathcal{P}(n)}
\log
\frac{\exp(\langle f_n,f_p\rangle/\tau_t)}
{\sum_{m \neq n}\exp(\langle f_n,f_m\rangle/\tau_t)},
\end{equation}
where $f_n=f_i(x_n)$ and $\tau_t$ is the task temperature. This objective preserves local semantic discrimination without requiring paired multimodal supervision.

\paragraph{Local structure construction.}
Using the adapted single-modality features, each student constructs batch-level group prototypes:
\begin{equation}
u_i^g=\frac{1}{|\mathcal{B}_i^g|}\sum_{n:y_n=g} f_n,
\end{equation}
where $\mathcal{B}_i^g$ denotes the set of batch samples belonging to group $g$. Based on $\{u_i^g\}$, the student builds a local affinity matrix $M_i$ using the same rule as on the teacher side. Importantly, this local graph is constructed in the exact same semantic group space as the teacher-side topology, facilitating direct compatibility with the global structural blueprint. The corresponding local support mask $A_i$ is defined by $(A_i)_{gh}=1$ only when both groups $g$ and $h$ are observed on student client $i$.

\paragraph{Topology-aware structural transfer.}
FedTaste aligns the local relational structure of student $i$ to the global blueprint only on edges that are simultaneously reliable, globally active, and locally supported:
\begin{equation}
\mathcal{L}_{\text{topo}}^{(i)}=
\frac{\left\| (S \odot A_{\text{g}} \odot A_i)\odot (M_i-\hat{M}_{\text{g}}) \right\|_F^2}
{\max\!\left(1,\left\|S \odot A_{\text{g}} \odot A_i\right\|_0\right)}.
\end{equation}
By restricting transfer to high-confidence relations, FedTaste encourages the student to absorb only globally verified semantic structure, mitigating the risk of incorporating unverified structural noise from isolated local distributions.

\paragraph{Spectral consistency and parameter regularization.}
Edge-level alignment alone may not sufficiently constrain the overall spectral properties of the student graph. To further stabilize adaptation, we regularize the low-frequency structure of $M_i$. Let $L_i$ be the local normalized Laplacian, and let $\Lambda_i=[\lambda_2^{(i)},\dots,\lambda_{K+1}^{(i)}]^\top$ collect its $K$ smallest non-trivial eigenvalues. The spectral loss is
\begin{equation}
\mathcal{L}_{\text{spec}}^{(i)}=\|\Lambda_i-\Lambda_{\text{g}}\|_2^2.
\end{equation}
By matching these principal eigenvalues derived from the normalized Laplacian, FedTaste aligns the macro-cluster structures, enhancing practical stability and cross-client comparability despite varying local supports and sparsity patterns.

To prevent the prompts from introducing overly aggressive local perturbations, we apply a capacity constraint via Frobenius norm regularization:
\begin{equation}
\mathcal{L}_{\text{reg}}^{(i)}=\sum_d \|P_i^{(d)}\|_F^2.
\end{equation}
This penalty ensures that the prompts act as gentle spatial calibrators rather than dominating the feature extraction process, preventing the model from overfitting to the restricted single-modality view. The complete student objective is
\begin{equation}
\mathcal{L}_{\text{total}}^{(i)}=
\mathcal{L}_{\text{task}}^{(i)}+
\beta_r \mathcal{L}_{\text{topo}}^{(i)}+
\gamma \mathcal{L}_{\text{spec}}^{(i)}+
\eta \mathcal{L}_{\text{reg}}^{(i)},
\end{equation}
where the topology weight is warmed up as $\beta_r=\beta \cdot \min(1,r/R_w)$ with communication round $r$ to avoid injecting structural supervision prematurely.

\subsection{Asymmetric Federated Optimization}
\label{subsec:federated_optimization}

FedTaste follows a role-asymmetric yet parameter-efficient optimization scheme. In each communication round, both teacher and student clients keep the shared backbone frozen and update only the lightweight MASP parameters attached to their active branches. The asymmetry lies in their structural contribution rather than backbone updating: full-modality teacher clients extract and upload multimodal structural payloads $\Pi_k$, whereas missing-modality student clients receive the global structural blueprint $\mathcal{B}$ and adapt their partial representations accordingly. 

During aggregation, the server aggregates the corresponding branch-specific MASP parameter subsets from participating clients while consolidating teacher-side structural payloads into the global blueprint. This decoupling is particularly advantageous for resource-constrained edge devices, as it significantly minimizes both the local memory footprint and uplink bandwidth needed for synchronization. 

Although a dense affinity graph has a worst-case complexity of $\mathcal{O}(G^2)$, each client instantiates only the subset of semantic groups supported by its local partition. Consequently, the effective graph size is substantially smaller than the global retrieval universe. Unsupported edges are explicitly masked, and the server further sparsifies the global blueprint through confidence filtering and per-row Top-$K_n$ retention, cleanly bypassing the scalability bottlenecks of conventional relation matching.

\begin{table}[t]
\centering
\renewcommand{\arraystretch}{1.2}
\setlength{\tabcolsep}{4pt}
\caption{Statistics of evaluated datasets.}
\label{tab:datasets}
\resizebox{0.95\columnwidth}{!}{%
\begin{tabular}{cccccc}
\toprule
\multirow{2}{*}{Dataset} & \multicolumn{2}{c}{Images} & \multicolumn{2}{c}{Texts} & \multirow{2}{*}{Local Task} \\ 
\cmidrule(lr){2-3} \cmidrule(lr){4-5}
 & Train & Test & Train & Test &  \\ 
\midrule
Flickr30k & 29,000 & 1,000 & 145,000 & 5,000 & Retrieval \\
MS-COCO & 113,287 & 5,000 & 566,435 & 25,000 & Retrieval \\
\midrule
OrganCMNIST & 12,975 & 8,216 & -- & -- & Classification \\
Medical Abstr. & -- & -- & 11,550 & 2,888 & Classification \\
\bottomrule
\end{tabular}%
}
\end{table}

% ==================================================================================
% Main Results: Varying Heterogeneity and Participation
% ==================================================================================
\begin{table*}[t]
    \centering
    \caption{Performance under varying heterogeneity and participation. The evaluation metric is the sum of top-1 recalls.}
    \label{tab:breakdown_part1}
    \renewcommand{\arraystretch}{1.15}
    \normalsize
    \setlength{\aboverulesep}{0pt}
    \setlength{\belowrulesep}{0pt}
    \setlength{\extrarowheight}{0pt}
    \begin{tabularx}{0.9\linewidth}{
        >{\raggedright\arraybackslash\hsize=1.35\hsize}X
        *{6}{>{\centering\arraybackslash\hsize=0.78\hsize}X}
    }
        \toprule
        \multirow{3}{*}{Method}
        & \multicolumn{2}{c}{Default}
        & \multicolumn{2}{c}{More Heterogeneity}
        & \multicolumn{2}{c}{Less Participation} \\
        & \multicolumn{2}{c}{$\alpha=0.5,\ r=0.25$}
        & \multicolumn{2}{c}{$\alpha=0.1,\ r=0.25$}
        & \multicolumn{2}{c}{$\alpha=0.5,\ r=0.125$} \\
        \cmidrule(lr){2-3} \cmidrule(lr){4-5} \cmidrule(lr){6-7}
        & \textsc{Flickr} & \textsc{COCO}
        & \textsc{Flickr} & \textsc{COCO}
        & \textsc{Flickr} & \textsc{COCO} \\
        \midrule
        FedAvg (2017)  & 81.08 & 95.42 & 81.70 & 95.32 & 64.82 & 83.91 \\
        FedProx (2020) & 78.55 & 95.16 & 76.33 & 95.62 & 63.33 & 79.88 \\
        FedIoT (2022)  & 85.51 & 98.40 & 83.28 & 95.89 & 61.94 & 80.65 \\
        CreamFL (2023) & 74.83 & 95.26 & 80.00 & 91.41 & 66.85 & 78.97 \\
        FedCola (2024) & 91.96 & 105.10 & 91.82 & 100.83 & 88.85 & 102.30 \\
        \rowcolor{mygray}
        \textbf{FedTaste (Ours)} & \textbf{99.64} & \textbf{114.31} & \textbf{97.82} & \textbf{109.14} & \textbf{94.68} & \textbf{110.26} \\
        \bottomrule
    \end{tabularx}
\end{table*}

% ==================================================================================
% Main Results: Skewed Modality Distributions
% ==================================================================================
\begin{table*}[t]
    \centering
    \caption{Performance under skewed modality distributions. The evaluation metric is the sum of top-1 recalls.}
    \label{tab:breakdown_part2}
    \renewcommand{\arraystretch}{1.15}
    \normalsize
    \setlength{\aboverulesep}{0pt}
    \setlength{\belowrulesep}{0pt}
    \setlength{\extrarowheight}{0pt}
    \begin{tabularx}{0.9\linewidth}{
        >{\raggedright\arraybackslash\hsize=1.35\hsize}X
        *{6}{>{\centering\arraybackslash\hsize=0.78\hsize}X}
    }
        \toprule
        \multirow{3}{*}{Method}
        & \multicolumn{2}{c}{More Image}
        & \multicolumn{2}{c}{More Text}
        & \multicolumn{2}{c}{Fewer Image-Text} \\
        & \multicolumn{2}{c}{$r=(0.33,\,0.25,\,0.25)$}
        & \multicolumn{2}{c}{$r=(0.25,\,0.33,\,0.25)$}
        & \multicolumn{2}{c}{$r=(0.25,\,0.25,\,0.125)$} \\
        \cmidrule(lr){2-3} \cmidrule(lr){4-5} \cmidrule(lr){6-7}
        & \textsc{Flickr} & \textsc{COCO}
        & \textsc{Flickr} & \textsc{COCO}
        & \textsc{Flickr} & \textsc{COCO} \\
        \midrule
        FedAvg (2017)  & 78.28 & 97.28 & 79.69 & 96.69 & 61.12 & 75.10 \\
        FedProx (2020) & 79.25 & 95.39 & 77.59 & 94.96 & 59.19 & 73.08 \\
        FedIoT (2022)  & 82.74 & 95.04 & 78.02 & 97.04 & 60.34 & 76.14 \\
        CreamFL (2023) & 80.31 & 93.65 & 80.75 & 92.81 & 57.12 & 74.69 \\
        FedCola (2024) & 91.24 & 104.22 & 90.10 & 100.96 & 85.68 & 94.40 \\
        \rowcolor{mygray}
        \textbf{FedTaste (Ours)} & \textbf{97.15} & \textbf{113.72} & \textbf{97.67} & \textbf{109.42} & \textbf{91.64} & \textbf{102.77} \\
        \bottomrule
    \end{tabularx}
\end{table*}

\section{Experiments}
\label{sec:exper}
\vspace{0.25em}
\subsection{Experiment Configuration}
\label{subsec:experset}
\noindent\textbf{Evaluation Protocol and Datasets.}
We construct a comprehensive evaluation protocol using standard splits of Flickr30k~\cite{plummer2015flickr30k} and MS-COCO~\cite{chen2015microsoft} for multimodal retrieval. By default, we simulate a severely heterogeneous federation of $N=32$ clients (8 full-modality teachers, 12 image-only students, and 12 text-only students). The data is partitioned via a Dirichlet process ($\alpha=0.5$) with a participation rate of $r=0.25$ per round. To rigorously assess robustness against domain shifts, we extend our evaluation to a medical scenario utilizing OrganCMNIST~\cite{yang2023medmnist} and Medical Abstracts~\cite{schopf2022evaluating} as out-of-distribution (OOD) datasets with limited in-domain full-modality teachers. This extreme scarcity setting explicitly tests whether methods relying on general-domain auxiliary data suffer from negative transfer compared to distilling native structural knowledge.

\noindent\textbf{Baselines.}
We benchmark against five representative methods across distinct paradigms: (1) classical optimization with zero-filling (FedAvg~\cite{mcmahan2017communication}, FedProx~\cite{li2020federated}); (2) modality-aware aggregation (FedIoT~\cite{zhao2022multimodal}); (3) auxiliary-data-driven transfer (CreamFL~\cite{yu2023multimodal}); and (4) state-of-the-art first-order coordinate alignment (FedCola~\cite{sun2024towards}). For fair evaluation, all methods share identical data partitions, communication rounds, and local epochs. While baselines adopt their original recommended architectures, FedTaste operates under a strictly parameter-efficient regime, proving the efficacy of second-order topology alignment by updating only the lightweight MASPs.

\noindent\textbf{Implementation Details.}
All experiments are conducted on a server equipped with an NVIDIA H100 GPU. To leverage strong semantic priors while maintaining strict parameter efficiency, FedTaste uniquely employs the frozen visual and text encoders of CLIP (ViT-B/32), which naturally project inputs into a shared $D=512$ latent space. Consequently, only the lightweight Modality-Adaptive Structural Prompts (MASP) are updated locally. We use the AdamW optimizer (learning rate $2 \times 10^{-4}$, weight decay 0.01, batch size 64) for 50 communication rounds, with 5 local epochs per selected client. For the topology distillation and structural alignment, we retain $K=5$ non-trivial eigenvalues and set the baseline topology weight to $\beta=0.5$ with a warm-up horizon of $R_w=10$ rounds. Further exhaustive details (e.g., baseline configurations, exact MASP insertion layers, and per-round communication breakdowns) are deferred to the supplementary material.

\subsection{Main Performance Comparison}
\label{subsec:main_results}

We evaluate FedTaste against state-of-the-art baselines across six federated scenarios. As shown in Tables~\ref{tab:breakdown_part1} and \ref{tab:breakdown_part2}, FedTaste consistently achieves the best performance on both Flickr30k and MS-COCO. These results highlight the effectiveness of transferring group-level semantic structure to address modality missingness, outperforming existing generative, external-data-driven, and first-order alignment paradigms.

\vspace{1mm}\noindent\textbf{Performance under Varying Heterogeneity and Participation.}
Table~\ref{tab:breakdown_part1} reports the performance under varying statistical conditions. Under the default setting ($\alpha=0.5$), FedTaste reaches an R@1$_{\text{sum}}$ of 99.64 on Flickr30k, outperforming the strongest baseline, FedCola (91.96). When data heterogeneity is exacerbated ($\alpha=0.1$), FedTaste preserves its significant lead, showing that group-level semantic topology remains robust against severe representation drift. Furthermore, under sparse client participation ($r=0.125$), FedTaste maintains a clear advantage (e.g., 94.68 vs. 88.85 for FedCola on Flickr30k), indicating that the aggregated global blueprint provides stable structural guidance even with infrequent synchronization.

\vspace{1mm}\noindent\textbf{Robustness to Skewed Modality Distributions.}
Table~\ref{tab:breakdown_part2} evaluates the models under three skewed modality distributions, which alter the ratio of full-modality teachers to missing-modality students. Despite the constrained availability of cross-modal structural knowledge, FedTaste consistently outperforms the baselines. This performance gap is particularly evident in the Fewer Image-Text scenario, where full-modality anchor clients are strictly limited. While methods relying on absolute coordinates or auxiliary datasets suffer severe performance drops under such scarcity, FedTaste effectively distills and propagates a stable blueprint from limited full-modality clients, highlighting its reliability in federations with severe teacher scarcity.

% ==================================================================================
% D. System Efficiency Analysis
% ==================================================================================
\subsection{Communication and Efficiency}
\label{subsec:efficiency}

Table~\ref{tab:efficiency} compares the retrieval performance and communication overhead of FedTaste against representative baselines. 
FedTaste achieves a superior accuracy-efficiency trade-off, demonstrating that its performance gains are not obtained at the expense of inflated communication costs.
This efficiency stems fundamentally from our asymmetric federated optimization strategy. 
Teacher clients transmit their MASP updates alongside a support-aware sparse topological payload, strictly bounded by $\mathcal{O}(G^2)$. 
Meanwhile, missing-modality student clients only synchronize the lightweight Modality-Adaptive Structural Prompts (MASP) associated with their active branches. 
As a result, the structural transfer introduces minimal overhead, substantially reducing the communication burden on resource-constrained edge devices while ensuring effective cross-modal collaboration.

\begin{table}[t]
    \centering
    \caption{Communication cost and retrieval performance comparison on Flickr30k.}
    \label{tab:efficiency}
    % 1. 样式设置
    \renewcommand{\arraystretch}{1.1} % 增加行高
    \small % 字体大小
    \setlength{\aboverulesep}{0pt}
    \setlength{\belowrulesep}{0pt}
    \setlength{\extrarowheight}{.75ex}
    \begin{tabularx}{0.95\linewidth}{ l >{\hsize=1.2\hsize}Y >{\hsize=0.8\hsize}Y }
        \toprule
        Method & Comm. Cost (MB) & R@1$_{\text{sum}}$ \\
        \midrule
        FedAvg  & 208.81 & 81.08 \\
        FedProx & 209.12 & 78.55 \\
        CreamFL & 211.74 & 74.83 \\
        FedIoT  & 219.45 & 85.51 \\
        FedCola & 371.26 & 91.96 \\
        % 4. 添加灰色背景
        \rowcolor{mygray} 
        \textbf{FedTaste} & \textbf{174.58} & \textbf{99.64} \\
        \bottomrule
    \end{tabularx}
    \vspace{-0.5em}
\end{table}

\begin{table}[t]
    \centering
    \caption{Impact of domain shift on retrieval performance (Sum of R@1). \textbf{Ideal} and \textbf{Noisy} denote in-domain and out-of-distribution external data, respectively. \textbf{$\Delta$} is the relative performance drop.}
    \label{tab:domain_shift}
    \renewcommand{\arraystretch}{1.1}
    \small
    \setlength{\aboverulesep}{0pt}
    \setlength{\belowrulesep}{0pt}
    \setlength{\extrarowheight}{.75ex}
    \begin{tabularx}{0.95\linewidth}{>{\raggedright\arraybackslash}X >{\centering\arraybackslash}X >{\centering\arraybackslash}X >{\centering\arraybackslash}X}
        \toprule
        Method & Ideal & Noisy & \textbf{$\Delta$} \\
        \midrule
        CreamFL & 74.83 & 48.54 & -35.13\% \\
        FedIoT & 85.51 & 82.05 & -4.05\% \\
        FedCola & 91.96 & 75.82 & -17.55\% \\
        \rowcolor{mygray}
        \textbf{FedTaste} & \textbf{99.64} & \textbf{97.55} & \textbf{-2.10\%} \\
        \bottomrule
    \end{tabularx}
    \vspace{-0.5em}
\end{table}
% ==================================================================================
% E. Robustness Analysis
% ==================================================================================
\subsection{Robustness under Domain Shift}
\label{subsec:robustness}

We evaluate the robustness of FedTaste against domain shifts, a critical challenge in real-world deployments. To simulate this scenario, we construct a specialized medical federation using the OrganCMNIST and Medical Abstracts datasets. Under this setting, baseline methods relying on external auxiliary datasets (e.g., general-domain image-text pairs) face a significant domain gap. This semantic mismatch can trigger negative transfer and degrade the alignment performance. 

In contrast, FedTaste entirely bypasses the reliance on exogenous data. It distills structural knowledge exclusively from the limited full-modality clients natively present within the target federation. By propagating a strictly in-domain global structural blueprint, FedTaste ensures the transferred topology matches the target distribution, thereby maintaining stable performance and effectively mitigating the risk of negative transfer.

\begin{figure}[t]
    \centering
    \includegraphics[width=0.95\columnwidth]{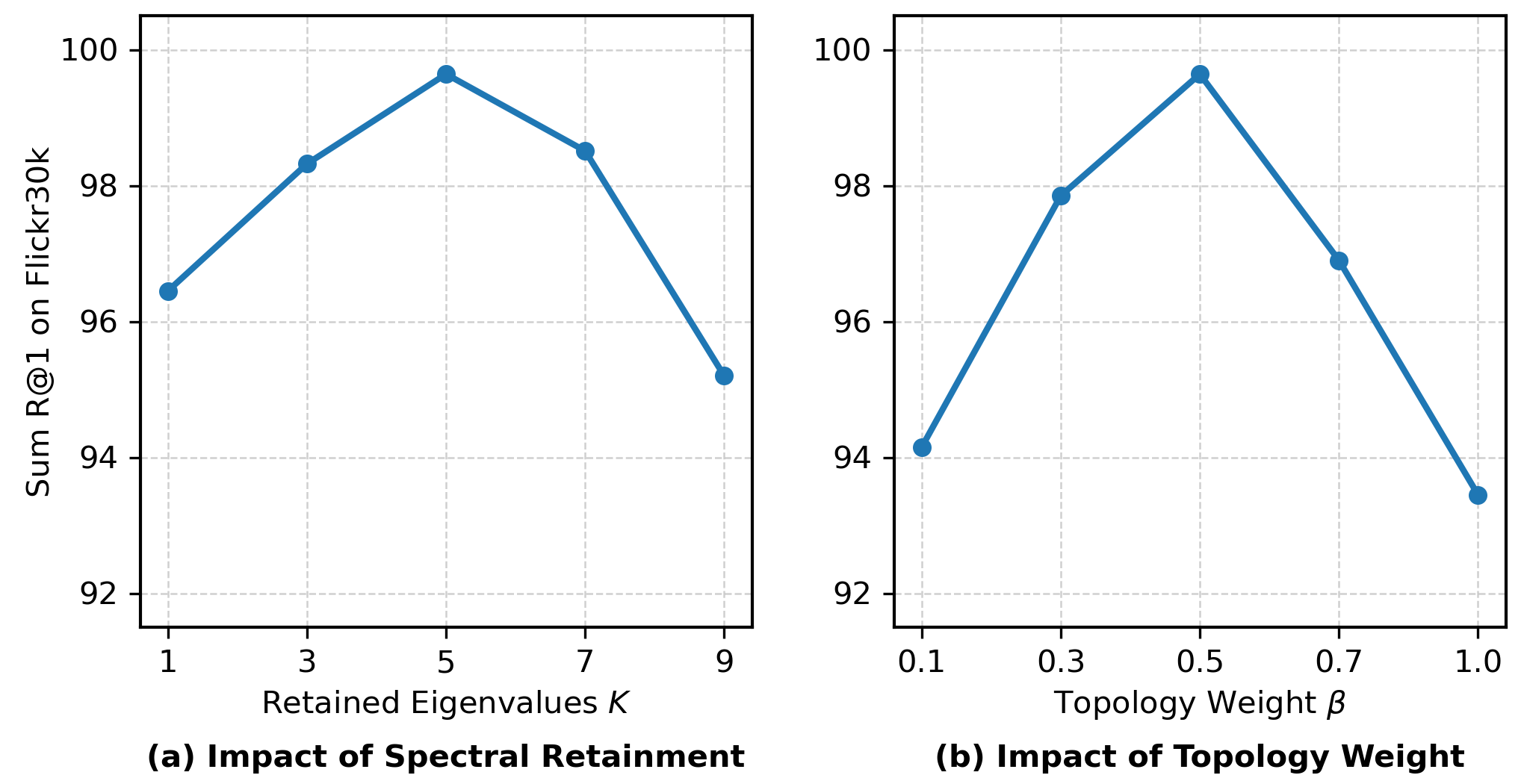}
    \caption{Hyperparameter sensitivity analysis of the retained eigenvalues $K$ and the topology weight $\beta$.}
    \label{fig:hparam_sensitivity}
    \vspace{-0.5em}
\end{figure}

% ==================================================================================
% H. Hyperparameter Sensitivity Analysis
% ==================================================================================
\subsection{Hyperparameter Sensitivity Analysis}
\label{subsec:hyperparameter}

We investigate the sensitivity of FedTaste to two key hyperparameters: the number of retained eigenvalues $K$ for spectral consistency, and the topology weight $\beta$.

\vspace{1mm}\noindent\textbf{Impact of Spectral Retainment ($K$).}
The parameter $K$ controls the granularity of the global topology preserved during spectral alignment. Thanks to the symmetric normalized Laplacian which bounds the spectral range, adjusting $K$ reliably focuses the alignment on macro-level structures. As shown in Figure~\ref{fig:hparam_sensitivity}(a), when $K$ is too small (e.g., $K=1$), the retained low-frequency information is insufficient to provide meaningful structural guidance. Conversely, an overly large $K$ (e.g., $K=9$) introduces unstable, high-frequency local noise. The peak performance at $K=5$ indicates an optimal balance between structural transfer and optimization stability.

\vspace{1mm}\noindent\textbf{Impact of Topology Weight ($\beta$).}
The weight $\beta$ determines the strength of structural transfer during local adaptation. Figure~\ref{fig:hparam_sensitivity}(b) demonstrates a clear inverted-U trend. A small weight ($\beta=0.1$) provides weak supervision, failing to adequately bridge the modality gap. However, an excessively large weight ($\beta=1.0$) forces the topology term to overwhelm the local task objective, severely degrading discriminative learning. The moderate value of $\beta=0.5$ yields the best performance, indicating that structural alignment must complement rather than suppress the task-specific gradients.

\subsection{Ablation Study}
\label{subsec:ablation}

We conduct ablation studies on the Flickr30k dataset to validate the core design components of FedTaste: the individual contribution of each module and the advantage of structural transfer over first-order alignment.

\vspace{1mm}\noindent\textbf{Effectiveness of Key Components.}
Table~\ref{tab:unified_ablation} presents a progressive evaluation of FedTaste under the default heterogeneous federated setting. Starting from the baseline, integrating Modality-Adaptive Structural Prompts (MASP) yields an initial performance gain by enabling lightweight local adaptation. However, the most significant improvement stems from the topology-aware structural transfer, which explicitly aligns partial representations of missing-modality clients with the teacher-derived global blueprint. Furthermore, spectral consistency regularization stabilizes the macro-level local geometry against data skew, and parameter regularization prevents excessive local perturbations. The complete FedTaste model achieves the best overall performance, demonstrating that the synergy between lightweight prompt adaptation and structural calibration is essential for effective cross-modal alignment.

\vspace{1mm}\noindent\textbf{Structural Transfer vs. First-Order Alignment.}
To isolate the source of improvement, we replace our structural transfer mechanism with a direct first-order prototype alignment variant (see Table~\ref{tab:unified_ablation}). The results indicate that aligning absolute feature coordinates is substantially less effective. Under severe Non-IID conditions, local embedding spaces drift across clients, rendering absolute coordinate matching fragile. Conversely, our structural transfer targets relative semantic relations, which remain consistent across disparate clients and modalities. This confirms that FedTaste's superiority originates fundamentally from its stable second-order transfer target rather than relying solely on parameter-efficient tuning.

\begin{table}[t]
    \centering
    \caption{Unified ablation study of FedTaste on the Flickr30k dataset. All results are reported in terms of Sum R@1.}
    \label{tab:unified_ablation}
    \renewcommand{\arraystretch}{1.1}
    \small
    \setlength{\aboverulesep}{0pt}
    \setlength{\belowrulesep}{0pt}
    \setlength{\extrarowheight}{.75ex}
    \begin{tabularx}{0.95\linewidth}{
        >{\hsize=0.80\hsize\centering\arraybackslash}X
        >{\hsize=0.9\hsize\centering\arraybackslash}X
        >{\hsize=1.30\hsize\centering\arraybackslash}X
        >{\hsize=0.9\hsize\centering\arraybackslash}X
        >{\hsize=0.9\hsize\centering\arraybackslash}X
        >{\hsize=1.0\hsize\centering\arraybackslash}X
        >{\hsize=1.2\hsize\centering\arraybackslash}X
    }
        \toprule
        Model & MASP & 1st-Ord. & \textbf{$\mathcal{L}_{\text{topo}}$} & \textbf{$\mathcal{L}_{\text{spec}}$} & \textbf{$\mathcal{L}_{\text{reg}}$} & \textbf{$R@1_{\text{sum}}$} \\
        \midrule
        A & \ding{55} & \ding{55} & \ding{55} & \ding{55} & \ding{55} & 81.08 \\
        B & \ding{51} & \ding{55} & \ding{55} & \ding{55} & \ding{55} & 90.84 \\
        C & \ding{51} & \ding{51} & \ding{55} & \ding{55} & \ding{55} & 93.55 \\
        D & \ding{51} & \ding{55} & \ding{51} & \ding{55} & \ding{55} & 96.18 \\
        E & \ding{51} & \ding{55} & \ding{51} & \ding{51} & \ding{55} & 98.37 \\
        \rowcolor{mygray}
        \textbf{F} & \textbf{\ding{51}} & \textbf{\ding{55}} & \textbf{\ding{51}} & \textbf{\ding{51}} & \textbf{\ding{51}} & \textbf{99.64} \\
        \bottomrule
    \end{tabularx}
\end{table}

% ==================================================================================
% F. Qualitative Visualization
% ==================================================================================
\subsection{Qualitative Visualization}
\label{subsec:visualization}

To qualitatively validate the efficacy of our method, we visualize the learned representation space through cross-modal affinity heatmaps and t-SNE projections.

\begin{figure}[t]
    \centering
    \includegraphics[width=0.97\columnwidth]{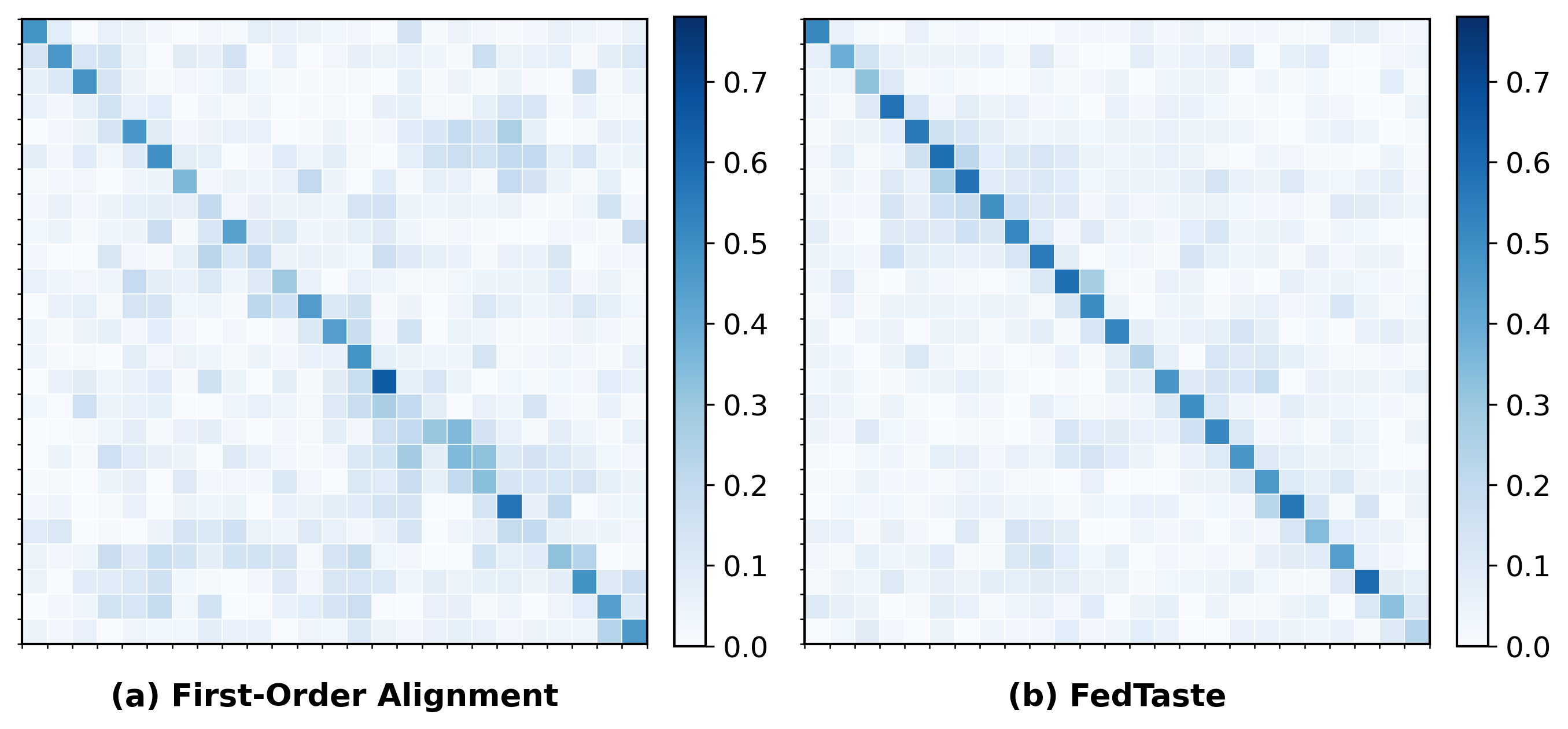}
    \vspace{-0.1cm}
    \caption{Cross-modal affinity heatmaps under first-order prototype alignment and FedTaste.}
    \label{fig:heatmap}
    \vspace{-0.3cm}
\end{figure}

\noindent\textbf{Cross-Modal Affinity Heatmap.}
As shown in Figure~\ref{fig:heatmap}, compared with naive first-order baselines, FedTaste exhibits a more concentrated diagonal pattern in the cross-modal affinity map. This provides clear evidence that anchoring local updates to the global structural blueprint effectively mitigates cross-client and cross-modal drift. It promotes tighter alignment for semantically matched associations while maintaining the discriminative sparsity of unmatched off-diagonal responses.

\begin{figure}[t]
    \centering
    \includegraphics[width=0.96\columnwidth]{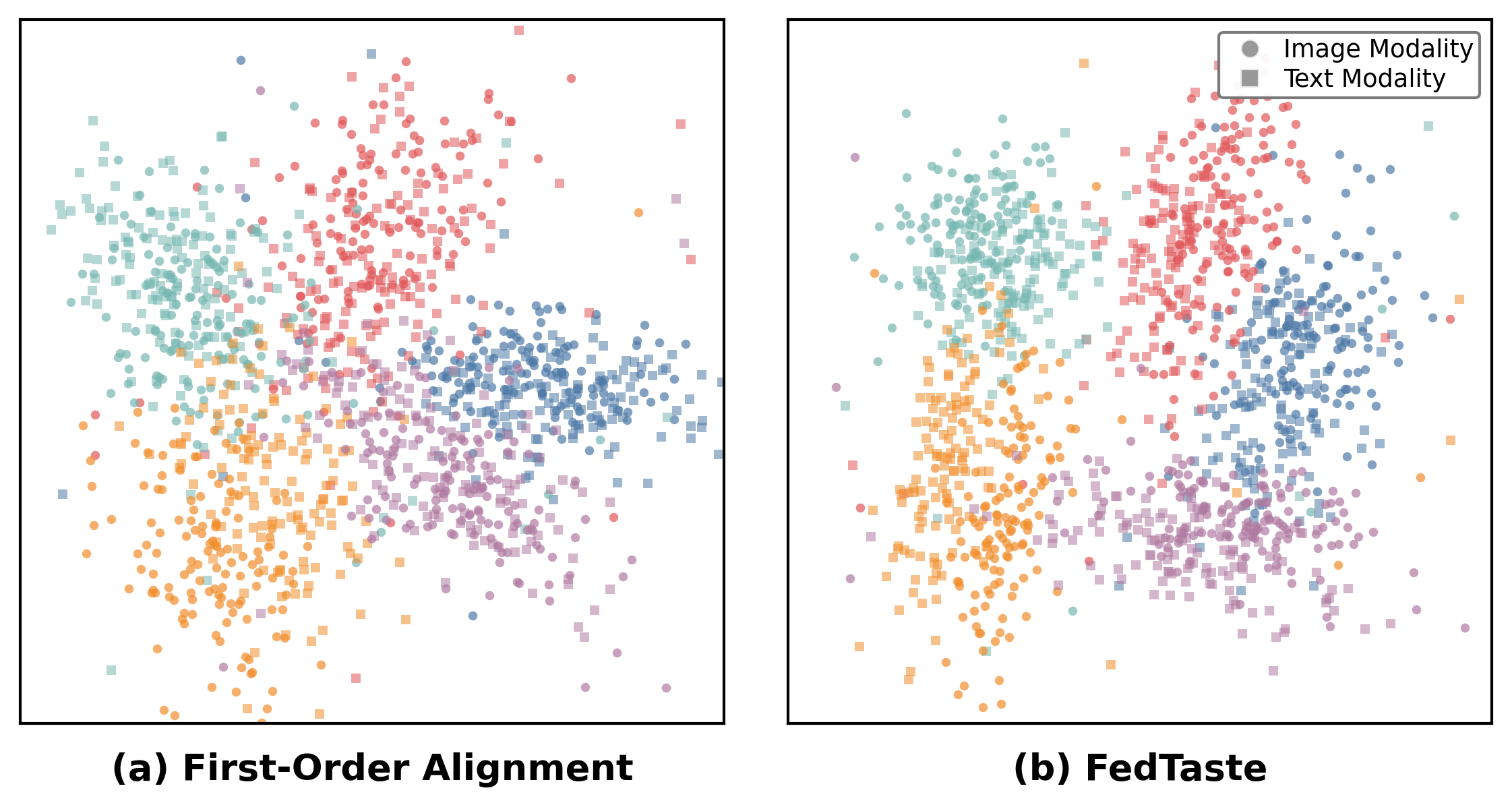}
    \vspace{-0.1cm}
    \caption{t-SNE visualization of latent representations under first-order alignment and FedTaste on selected semantic groups.}
    \label{fig:tsne}
    \vspace{-0.3cm}
\end{figure}

\noindent\textbf{t-SNE Projection of the Latent Space.}
As illustrated in Figure~\ref{fig:tsne}, in the projected latent space, FedTaste produces coherent and better-aligned clusters for corresponding image and text samples from selected semantic groups. Unlike point-wise alignment methods that easily collapse under severe Non-IID heterogeneity, this clustering pattern indicates that transferring shared group-level semantic topology effectively regularizes local representation spaces, thereby yielding a more robust and unified cross-modal geometry.

\section{Conclusion}
\label{sec:conclusion}

In this paper, we propose FedTaste, a topology-aware structural transfer framework for multimodal federated learning under arbitrary modality missingness. Bypassing fragile first-order coordinate alignment and costly generative imputation, FedTaste transfers stable group-level semantic topology from full-modality teachers to missing-modality students. This is achieved via a confidence-aware global structural blueprint, lightweight Modality-Adaptive Structural Prompts, and asymmetric federated optimization. By anchoring local adaptations to a shared relational graph, our framework efficiently calibrates partial representations without requiring full-backbone synchronization or privacy-compromising auxiliary datasets. Extensive experiments on standard cross-modal benchmarks validate that FedTaste achieves a superior accuracy-efficiency trade-off, demonstrating remarkable robustness under severe Non-IID conditions, skewed modality distributions, and extreme teacher scarcity, ultimately paving the way for scalable and privacy-preserving multimodal systems at the network edge. Future work will extend this paradigm to highly dynamic federated environments with asynchronous participation, and generalize our strategy to broader heterogeneous modalities (e.g., audio and sensor data) for IoT applications.
%%
%% The next two lines define the bibliography style to be used, and
%% the bibliography file.
\bibliographystyle{ACM-Reference-Format}
\bibliography{sample-base}

@String{Computing = "Computing" }

@String{Computer = "{IEEE} Computer" }

@String{Springer = "Springer-Verlag" }

@article{che2023multimodal,
  title={Multimodal federated learning: A survey},
  author={Che, Liwei and Wang, Jiaqi and Zhou, Yao and Ma, Fenglong},
  journal={Sensors},
  volume={23},
  number={15},
  pages={6986},
  year={2023},
  publisher={MDPI}
}

@article{li2021privacy,
  title={Privacy-preserved federated learning for autonomous driving},
  author={Li, Yijing and Tao, Xiaofeng and Zhang, Xuefei and Liu, Junjie and Xu, Jin},
  journal={IEEE Transactions on Intelligent Transportation Systems},
  volume={23},
  number={7},
  pages={8423--8434},
  year={2021},
  publisher={IEEE}
}

@article{nguyen2022federated,
  title={Federated learning for smart healthcare: A survey},
  author={Nguyen, Dinh C and Pham, Quoc-Viet and Pathirana, Pubudu N and Ding, Ming and Seneviratne, Aruna and Lin, Zihuai and Dobre, Octavia and Hwang, Won-Joo},
  journal={ACM Computing Surveys (Csur)},
  volume={55},
  number={3},
  pages={1--37},
  year={2022},
  publisher={ACM New York, NY}
}

@inproceedings{zong2021fedcmr,
  title={FedCMR: Federated cross-modal retrieval},
  author={Zong, Linlin and Xie, Qiujie and Zhou, Jiahui and Wu, Peiran and Zhang, Xianchao and Xu, Bo},
  booktitle={Proceedings of the 44th International ACM SIGIR Conference on Research and Development in Information Retrieval},
  pages={1672--1676},
  year={2021}
}

@article{huang2024multimodal,
  title={Multimodal federated learning: Concept, methods, applications and future directions},
  author={Huang, Wei and Wang, Dexian and Ouyang, Xiaocao and Wan, Jihong and Liu, Jia and Li, Tianrui},
  journal={Information Fusion},
  volume={112},
  pages={102576},
  year={2024},
  publisher={Elsevier}
}

@article{song2024tackling,
  title={Tackling modality-heterogeneous client drift holistically for heterogeneous multimodal federated learning},
  author={Song, Haoyue and Wang, Jiacheng and Zhou, Jianjun and Wang, Liansheng},
  journal={IEEE Transactions on Medical Imaging},
  volume={44},
  number={4},
  pages={1931--1941},
  year={2024},
  publisher={IEEE}
}

@inproceedings{sun2024towards,
  title={Towards multi-modal transformers in federated learning},
  author={Sun, Guangyu and Mendieta, Matias and Dutta, Aritra and Li, Xin and Chen, Chen},
  booktitle={European Conference on Computer Vision},
  pages={229--246},
  year={2024},
  organization={Springer}
}

@article{cao2026heterogeneous,
  title={Heterogeneous Multimodal Federated Learning with Missing Modality via Mask-Restoration and Self-Guidance},
  author={Cao, Zhibo and Hao, Kuangrong and Hao, Lingguang and Wei, Bing and Ren, Lihong},
  journal={IEEE Transactions on Multimedia},
  year={2026},
  publisher={IEEE}
}

@article{yu2023multimodal,
  title={Multimodal federated learning via contrastive representation ensemble},
  author={Yu, Qiying and Liu, Yang and Wang, Yimu and Xu, Ke and Liu, Jingjing},
  journal={arXiv preprint arXiv:2302.08888},
  year={2023}
}

@article{liang2026fedcoop,
  title={FedCoop: Co-optimized Latent Diffusion for Asymmetric Multimodal Federated Learning},
  author={Liang, Haochen and Chen, Chaomeng and Yang, Jing and Ma, Hui and Cao, Junzhe and Yu, Zitong},
  journal={Authorea Preprints},
  year={2026},
  publisher={Authorea}
}

@inproceedings{chen2024fedmbridge,
  title={FedMBridge: Bridgeable multimodal federated learning},
  author={Chen, Jiayi and Zhang, Aidong},
  booktitle={Forty-first International Conference on Machine Learning},
  year={2024}
}

@article{zhang2025unimodal,
  title={Unimodal training-multimodal prediction: Cross-modal federated learning with hierarchical aggregation},
  author={Zhang, Rongyu and Chi, Xiaowei and Zhang, Wenyi and Liu, Guiliang and Wang, Dan and Wang, Fangxin},
  journal={IEEE Transactions on Mobile Computing},
  year={2025},
  publisher={IEEE}
}

@inproceedings{mcmahan2017communication,
  title={Communication-efficient learning of deep networks from decentralized data},
  author={McMahan, Brendan and Moore, Eider and Ramage, Daniel and Hampson, Seth and y Arcas, Blaise Aguera},
  booktitle={Artificial intelligence and statistics},
  pages={1273--1282},
  year={2017},
  organization={Pmlr}
}

@article{li2020federated,
  title={Federated optimization in heterogeneous networks},
  author={Li, Tian and Sahu, Anit Kumar and Zaheer, Manzil and Sanjabi, Maziar and Talwalkar, Ameet and Smith, Virginia},
  journal={Proceedings of Machine learning and systems},
  volume={2},
  pages={429--450},
  year={2020}
}

@inproceedings{zhao2022multimodal,
  title={Multimodal federated learning on iot data},
  author={Zhao, Yuchen and Barnaghi, Payam and Haddadi, Hamed},
  booktitle={2022 IEEE/ACM seventh international conference on internet-of-things design and implementation (ioTDI)},
  pages={43--54},
  year={2022},
  organization={IEEE}
}

@inproceedings{radford2021learning,
  title={Learning transferable visual models from natural language supervision},
  author={Radford, Alec and Kim, Jong Wook and Hallacy, Chris and Ramesh, Aditya and Goh, Gabriel and Agarwal, Sandhini and Sastry, Girish and Askell, Amanda and Mishkin, Pamela and Clark, Jack and others},
  booktitle={International conference on machine learning},
  pages={8748--8763},
  year={2021},
  organization={PmLR}
}

@inproceedings{shi2024clip,
  title={Clip-guided federated learning on heterogeneity and long-tailed data},
  author={Shi, Jiangming and Zheng, Shanshan and Yin, Xiangbo and Lu, Yang and Xie, Yuan and Qu, Yanyun},
  booktitle={Proceedings of the AAAI Conference on Artificial Intelligence},
  volume={38},
  number={13},
  pages={14955--14963},
  year={2024}
}

@inproceedings{feng2023fedmultimodal,
  title={Fedmultimodal: A benchmark for multimodal federated learning},
  author={Feng, Tiantian and Bose, Digbalay and Zhang, Tuo and Hebbar, Rajat and Ramakrishna, Anil and Gupta, Rahul and Zhang, Mi and Avestimehr, Salman and Narayanan, Shrikanth},
  booktitle={Proceedings of the 29th ACM SIGKDD conference on knowledge discovery and data mining},
  pages={4035--4045},
  year={2023}
}

@article{tan2023fedsea,
  title={Fedsea: Federated learning via selective feature alignment for non-iid multimodal data},
  author={Tan, Min and Feng, Yinfu and Chu, Lingqiang and Shi, Jingcheng and Xiao, Rong and Tang, Haihong and Yu, Jun},
  journal={IEEE Transactions on Multimedia},
  volume={26},
  pages={5807--5822},
  year={2023},
  publisher={IEEE}
}

@inproceedings{jiang2022harmofl,
  title={Harmofl: Harmonizing local and global drifts in federated learning on heterogeneous medical images},
  author={Jiang, Meirui and Wang, Zirui and Dou, Qi},
  booktitle={Proceedings of the AAAI conference on artificial intelligence},
  volume={36},
  number={1},
  pages={1087--1095},
  year={2022}
}

@inproceedings{mendieta2022local,
  title={Local learning matters: Rethinking data heterogeneity in federated learning},
  author={Mendieta, Matias and Yang, Taojiannan and Wang, Pu and Lee, Minwoo and Ding, Zhengming and Chen, Chen},
  booktitle={Proceedings of the IEEE/CVF conference on computer vision and pattern recognition},
  pages={8397--8406},
  year={2022}
}

@article{wu2024topology,
  title={Topology-aware federated learning in edge computing: A comprehensive survey},
  author={Wu, Jiajun and Dong, Fan and Leung, Henry and Zhu, Zhuangdi and Zhou, Jiayu and Drew, Steve},
  journal={ACM Computing Surveys},
  volume={56},
  number={10},
  pages={1--41},
  year={2024},
  publisher={ACM New York, NY}
}

@inproceedings{hu2026fedtopo,
  title={FedTopo: Topology-Informed Representation Alignment in Federated Learning Under Non-IID Conditions},
  author={Hu, Ke and Xiang, Liyao and Tang, Peng and Qiu, Weidong},
  booktitle={Proceedings of the AAAI Conference on Artificial Intelligence},
  volume={40},
  number={26},
  pages={21849--21857},
  year={2026}
}

@article{chen2024enhancing,
  title={Enhancing decentralized and personalized federated learning with topology construction},
  author={Chen, Suo and Xu, Yang and Xu, Hongli and Ma, Zhenguo and Wang, Zhiyuan},
  journal={IEEE Transactions on Mobile Computing},
  volume={23},
  number={10},
  pages={9692--9707},
  year={2024},
  publisher={IEEE}
}

@inproceedings{ngo2025higda,
  title={Higda: Hierarchical graph of nodes to learn local-to-global topology for semi-supervised domain adaptation},
  author={Ngo, Ba Hung and Bui, Doanh C and Do-Tran, Nhat-Tuong and Choi, Tae Jong},
  booktitle={Proceedings of the AAAI conference on artificial intelligence},
  volume={39},
  number={6},
  pages={6191--6199},
  year={2025}
}

@article{yang2022geometric,
  title={Geometric knowledge distillation: Topology compression for graph neural networks},
  author={Yang, Chenxiao and Wu, Qitian and Yan, Junchi},
  journal={Advances in Neural Information Processing Systems},
  volume={35},
  pages={29761--29775},
  year={2022}
}

@inproceedings{kavalionak2021impact,
  title={Impact of network topology on the convergence of decentralized federated learning systems},
  author={Kavalionak, Hanna and Carlini, Emanuele and Dazzi, Patrizio and Ferrucci, Luca and Mordacchini, Matteo and Coppola, Massimo},
  booktitle={2021 IEEE Symposium on Computers and Communications (ISCC)},
  pages={1--6},
  year={2021},
  organization={IEEE}
}

@article{wei2024joint,
  title={Joint participant and learning topology selection for federated learning in edge clouds},
  author={Wei, Xinliang and Ye, Kejiang and Shi, Xinghua and Xu, Cheng-Zhong and Wang, Yu},
  journal={IEEE Transactions on Parallel and Distributed Systems},
  volume={35},
  number={8},
  pages={1456--1468},
  year={2024},
  publisher={IEEE}
}

@article{lin2023federated,
  title={Federated learning on multimodal data: A comprehensive survey},
  author={Lin, Yi-Ming and Gao, Yuan and Gong, Mao-Guo and Zhang, Si-Jia and Zhang, Yuan-Qiao and Li, Zhi-Yuan},
  journal={Machine Intelligence Research},
  volume={20},
  number={4},
  pages={539--553},
  year={2023},
  publisher={Springer}
}

@inproceedings{wang2024fedmmr,
  title={Fedmmr: Multi-modal federated learning via missing modality reconstruction},
  author={Wang, Shu and Qu, Zhe and Liu, Yuan and Kan, Shichao and Liang, Yixiong and Wang, Jianxin},
  booktitle={2024 IEEE International Conference on Multimedia and Expo (ICME)},
  pages={1--6},
  year={2024},
  organization={IEEE}
}

@inproceedings{plummer2015flickr30k,
  title={Flickr30k entities: Collecting region-to-phrase correspondences for richer image-to-sentence models},
  author={Plummer, Bryan A and Wang, Liwei and Cervantes, Chris M and Caicedo, Juan C and Hockenmaier, Julia and Lazebnik, Svetlana},
  booktitle={Proceedings of the IEEE international conference on computer vision},
  pages={2641--2649},
  year={2015}
}

@article{chen2015microsoft,
  title={Microsoft coco captions: Data collection and evaluation server},
  author={Chen, Xinlei and Fang, Hao and Lin, Tsung-Yi and Vedantam, Ramakrishna and Gupta, Saurabh and Doll{\'a}r, Piotr and Zitnick, C Lawrence},
  journal={arXiv preprint arXiv:1504.00325},
  year={2015}
}

@article{yang2023medmnist,
  title={Medmnist v2-a large-scale lightweight benchmark for 2d and 3d biomedical image classification},
  author={Yang, Jiancheng and Shi, Rui and Wei, Donglai and Liu, Zequan and Zhao, Lin and Ke, Bilian and Pfister, Hanspeter and Ni, Bingbing},
  journal={Scientific data},
  volume={10},
  number={1},
  pages={41},
  year={2023},
  publisher={Nature Publishing Group UK London}
}

@inproceedings{schopf2022evaluating,
  title={Evaluating unsupervised text classification: zero-shot and similarity-based approaches},
  author={Schopf, Tim and Braun, Daniel and Matthes, Florian},
  booktitle={Proceedings of the 2022 6th International Conference on Natural Language Processing and Information Retrieval},
  pages={6--15},
  year={2022}
}

@inproceedings{poudel2025multimodal,
  title={Multimodal federated learning with missing modalities through feature imputation network},
  author={Poudel, Pranav and Chhetri, Aavash and Gyawali, Prashnna and Leontidis, Georgios and Bhattarai, Binod},
  booktitle={Annual Conference on Medical Image Understanding and Analysis},
  pages={289--299},
  year={2025},
  organization={Springer}
}

@article{yan2025federated,
  title={Federated pseudo modality generation for incomplete multi-modal mri reconstruction},
  author={Yan, Yunlu and Feng, Chun-Mei and Li, Yuexiang and Li, Ping and Goh, Rick Siow Mong and Lei, Baiying and Wang, Weiming and Feng, David Dagan and Zhu, Lei},
  journal={IEEE Journal of Biomedical and Health Informatics},
  year={2025},
  publisher={IEEE}
}

@inproceedings{liu2025fedmobile,
  title={Fedmobile: Enabling knowledge contribution-aware multi-modal federated learning with incomplete modalities},
  author={Liu, Yi and Wang, Cong and Yuan, Xingliang},
  booktitle={Proceedings of the ACM on Web Conference 2025},
  pages={2775--2786},
  year={2025}
}

@article{yu2026robust,
  title={Robust multimodal federated learning for non-IID multimodal data with incompleteness},
  author={Yu, Songcan and Zhu, Kaiming and Liang, Feiyuan and Wang, Junbo and Kant, Krishna and Yin, Ling},
  journal={Future Generation Computer Systems},
  volume={174},
  pages={107948},
  year={2026},
  publisher={Elsevier}
}

@inproceedings{poudel2024car,
  title={Car-mfl: Cross-modal augmentation by retrieval for multimodal federated learning with missing modalities},
  author={Poudel, Pranav and Shrestha, Prashant and Amgain, Sanskar and Shrestha, Yash Raj and Gyawali, Prashnna and Bhattarai, Binod},
  booktitle={International Conference on Medical Image Computing and Computer-Assisted Intervention},
  pages={102--112},
  year={2024},
  organization={Springer}
}

@article{tan2026fedafd,
  title={FedAFD: Multimodal Federated Learning via Adversarial Fusion and Distillation},
  author={Tan, Min and Ma, Junchao and Feng, Yinfu and Ding, Jiajun and Pan, Wenwen and Han, Tingting and Zheng, Qian and Kuang, Zhenzhong and Yu, Zhou},
  journal={arXiv preprint arXiv:2603.04890},
  year={2026}
}

@inproceedings{chen2022towards,
  title={Towards optimal multi-modal federated learning on non-IID data with hierarchical gradient blending},
  author={Chen, Sijia and Li, Baochun},
  booktitle={IEEE INFOCOM 2022-IEEE conference on computer communications},
  pages={1469--1478},
  year={2022},
  organization={IEEE}
}

@article{bao2025robust,
  title={A Robust Odor Mixture Quantification Method Based on Active Sensing Using Both QCM Frequency Shifts and Resistance Changes of Multiple Harmonics},
  author={Bao, Ziteng and Aleixandre, Manuel and Hasegawa, Shoichi and Nakamoto, Takamichi},
  journal={IEEE Sensors Journal},
  year={2025},
  publisher={IEEE}
}

@article{feng2024robust,
  title={Robust privacy-preserving recommendation systems driven by multimodal federated learning},
  author={Feng, Chenyuan and Feng, Daquan and Huang, Guanxin and Liu, Zuozhu and Wang, Zhenzhong and Xia, Xiang-Gen},
  journal={IEEE Transactions on Neural Networks and Learning Systems},
  volume={36},
  number={5},
  pages={8896--8910},
  year={2024},
  publisher={IEEE}
}

@article{hsu2024federated,
  title={Federated learning using multi-modal sensors with heterogeneous privacy sensitivity levels},
  author={Hsu, Chih-fan and Li, Yi-chen and Tsai, Chung-chi and Wang, Jian-kai and Hsu, Cheng-hsin},
  journal={ACM Transactions on Multimedia Computing, Communications and Applications},
  volume={20},
  number={11},
  pages={1--27},
  year={2024},
  publisher={ACM New York, NY}
}

\end{document}